\documentclass[aps,preprint,epsfig,rotate]{revtex4}
\begin{document}
%\begin{doublespace}
\title{Numerical evaluations of the isotopic shifts and lowest-order QED corrections for the ground $1^1S-$states of the 
       ${}^{3}$He and ${}^{4}$He atoms.}

 \author{Alexei M. Frolov}
 \email[E--mail address: ]{afrolov@uwo.ca}

\affiliation{Department of Applied Mathematics \\
 University of Western Ontario, London, Ontario N6H 5B7, Canada}

\date{\today}

\begin{abstract}

The both mass-dependent and field shift components of the isotopic shifts and the lowest order QED corrections for the ground 
(singlet) $1^1S(L = 0)-$states of the ${}^{3}$He and ${}^{4}$He atoms are determined to high accuracy. For the same states we 
also evaluated the lowest-order QED corrections and the corresponding recoil (or finite mass) corrections. In our calculations 
we have used the new (corrected) formula for the recoil correction to the lowest-order QED correction which can be applied to 
atoms/ions with arbitrary nuclear charge $Q \ge 1$.  

\end{abstract}

\maketitle
\newpage

\section{Introduction}

The goal of this communication is to determine different components of the isotope shift and the lowest-order  Quantum-Electrodynamics 
corrections (or QED corrections, for short) for the actual helium-3 and helium-4 atoms, i.e. for the two isotopes of the helium atoms 
which have the finite nuclear masses. In general, highly accurate calculations of the lowest-order QED corrections in few-electron 
atomic systems can be performed only with the use of the relativistic wave functions. However, approximate numerical values of the 
lowest-order QED corrections can be determined with the non-relativistic wave functions as the solutions of the Schr\"{o}dinger equation 
\cite{FrFi} for the bound states $H \Psi = E \Psi$, where $E < 0$ and $H$ is the non-relativistic Hamiltonian of the helium atom 
\begin{equation}
 H = -\frac{\hbar^{2}}{2 m_{e}} \Bigl( \nabla^{2}_{1} + \nabla^{2}_{2} + \frac{m_e}{M_N} \nabla^{2}_{N} \Bigr) - \frac{Q e^2}{r_{32}} - 
 \frac{Q e^2}{r_{31}} + \frac{e^2}{r_{21}} \label{Ham}
\end{equation}
where $\nabla_{i} = \Bigl( \frac{\partial}{\partial x_{i}}, \frac{\partial}{\partial y_{i}}, \frac{\partial}{\partial z_{i}} \Bigr)$ and
$i = 1, 2, 3(= N)$, where the notation $N$ stands for the nucleus. In Eq.(\ref{Ham}) the notation $\hbar$ stands for the reduced Planck 
constant, i.e. $\hbar = \frac{h}{2 \pi}$, and $e$ is the elementary electric charge. Everywhere below in this study the particles 1 and 2 
mean the electrons, while the particle 3 (or $N$) is the atomic nucleus with the mass $M_N \gg m_e$. For light atoms it is very convenient 
to perform all bound state calculations in atomic units where $\hbar = 1, m_e = 1$ and $e = 1$. In these units the velocity of light in 
vacuum $c$ numerically coincides with the inverse value of the dimensionless fine structure constant, i.e. $c = \alpha^{-1}$, where 
$\alpha = \frac{e^2}{\hbar c} \approx$ 7.2973525698$\cdot 10^{-3}$ \cite{CRC}. In atomic units the same Hamiltonian, Eq.(\ref{Ham}), is 
written in the form
\begin{equation}
 H = -\frac12 \Bigl( \nabla^{2}_{1} + \nabla^{2}_{2} + \frac{1}{M_N} \nabla^{2}_{N} \Bigr) - \frac{Q}{r_{32}} - \frac{Q}{r_{31}} + 
 \frac{1}{r_{21}} \label{Ham1}
\end{equation}
where $M_N$ is expressed in the electron mass $m_e$.

In this study we also apply the non-relativistic wave functions to determine some properties of the two-electron helium atom(s). All wave 
functions used in this analysis have been obtained as the solutions of the (non-relativistic) Schr\"{o}dinger equation with the Hamiltonian, 
Eq.(\ref{Ham1}). It should be emphasized that this approach works, if (and only if) we can construct the non-relativistic variational wave 
functions of very high accuracy. For the ground $1^{1}S(L = 0)-$states of the ${}^{\infty}$He, ${}^{4}$He and ${}^{3}$He atoms the highly 
accurate wave functions are constructed in the form of the following variational expansion (see, e.g., \cite{Fro98})
\begin{eqnarray}
 \Psi = \Bigl( 1 + \hat{P}_{12} \Bigr) \sum_{i=1}^{N} C_{i} \exp(-\alpha_{i} r_{32} - \beta_{i} r_{31} - \gamma_{i} r_{21})  \label{exp1} 
\end{eqnarray}
which is called the exponential variational expansion in the relative coordinates $r_{32}, r_{31}$ and $r_{21}$. Each of these three relative 
coordinates is defined as the difference between the corresponding Cartesian coordinates of the two particles, e.g., $r_{ij} = \mid {\bf r}_i 
- {\bf r}_j \mid$. It follows from this definition that the relative coordinates $r_{32}, r_{31}$ and $r_{21}$ are translationally and 
rotationally invariant. The coefficients $C_i$ are the linear (or variational) parameters of the variational expansion, Eq.(\ref{exp1}), while 
the parameters $\alpha_{i}, \beta_{i}$ and $\gamma_{i}$ are the non-linear (or varied) parameters of this expansion. In general, the total 
energy of the ground state of the He atom depends upon the total number of basis functions $N$, Eq.(\ref{exp1}), used in calculations. 
The operator $\hat{P}_{12}$ is the permutation operator for two identical particles (electrons). 

\section{Isotopic shift(s)}

Differences of the corresponding (atomic) total energies $E({}^{\infty}$He), $E({}^{4}$He), $E({}^{3}$He) coincide with the non-relativistic 
isotopic shifts for the isotopes of the helium atom(s). In this study we used the following nuclear masses (expressed in the electron mass $m_e$): 
$M({}^{3}{\rm He})$ = 5495.885269 $m_e$ and $M({}^{4}{\rm He})$ = 7294.2995363 $m_e$   These masses were used in earlier calculations of the 
${}^{3}$He and ${}^{4}$He atoms and they are very close to the values obtained in the recent nuclear and high-energy experiments. Note that in 
highly accurate computations these numerical values of nuclear masses (as well as all other physical constants) can be considered as `exact'. The 
corresponding corrections can be evaluated at the next step by performing numerical calculations with the different values of the nuclear masses 
and other physical constants. The difference of the total non-relativistic energies of the ${}^{3}$He and ${}^{4}$He atoms
\begin{eqnarray}
          E({}^{3}{\rm He}) - E({}^{4}{\rm He}) \approx 1.473271 518440 868154 71 \cdot 10^{-4} \; \; \; a.u. 
\end{eqnarray}
has a great interest in applications, since its numerical value can be measured in simple atomic experiments. In atomic physics the non-relativistic 
isotope shifts for light atoms are represented as a few-term sums. Note that if the momentum of the nucleus $N$ is known (it is designated below as 
${\bf P}_N$), then the isotopic shift for an atom with the nuclear mass $M$ equals to the expectation value of the $\frac{1}{2 M} {\bf P}^{2}_N$ 
operator, which corresponds to the kinetic energy of the atomic nucleus. However, since the relative coordinates $r_{32}, r_{31}$ and $r_{21}$ are 
translationally invariant, we can always assume that we are working in the center-of-mass system and this center of mass does not move, i.e. ${\bf 
P}_N + {\bf p}_1 + \ldots + {\bf p}_K = 0$ for the $K-$electron atom. This can be re-written in the different form 
\begin{eqnarray}
 {\bf P}^{2}_N = \sum^{K}_{i=1} {\bf p}^{2}_{i} + \sum^{K}_{i=2 (i \ne j)} \sum^{K-1}_{j=1} {\bf p}_i \cdot {\bf p}_j 
\end{eqnarray}
for the expectation values we have
\begin{eqnarray}
 \langle {\bf P}^{2}_N \rangle = K \langle {\bf p}^{2}_{1} \rangle + \frac{2 K (K - 1)}{2} \langle {\bf p}_1 \cdot {\bf p}_2 \rangle \label{ish1ft} 
\end{eqnarray}
where we used the fact that all electrons are identical particles. It follows from here that 
\begin{eqnarray}
 \frac{1}{K M} {\bf P}^{2}_N = \frac{1}{M} \langle {\bf p}^{2}_{1} \rangle + \frac{(K-1)}{M} \langle {\bf p}_1 \cdot {\bf p}_2 \rangle \label{ishift1} 
\end{eqnarray}
In particular, for two-electron atomic systems $K = 2$ one finds
\begin{eqnarray}
 \frac{1}{2 M} \langle {\bf P}^{2}_N \rangle = \frac{1}{M} \langle {\bf p}^{2}_{1} \rangle + \frac{1}{M} \langle {\bf p}_1 \cdot {\bf p}_2 \rangle 
 \label{ishift2} 
\end{eqnarray}
In other words, to determine the isotopic shift in the two-electron atoms/ions one neeeds to determine the expectation values of the ${\bf p}^{2}_{1}$
and ${\bf p}_1 \cdot {\bf p}_2$ operators. Note that the first operator is one-electron, while the second operator is a two-electron operator \cite{MQW}.
The first operator in Eq.(\ref{ishift2}) represents the normal mass shift, while the second operator represents the specific mass shift (for more details, 
see, Chapter 8 in \cite{FrFi}). As follows from Eq.(\ref{ishift2}) both of these components of the mass shift are mass dependent (or $M-$dependent). 

Thus, to determine mass-dependent components of the isotopic shifts one needs either to perform the direct calculations of the corresponding total energies
$E({}^{3}$He), $E({}^{4}$He) and $E({}^{\infty}$He) (as we did in this study), or to determine the expectation value of the ${\bf P}^{2}_N$ operator and/or
to evaluate to very high accuracy the $\langle {\bf p}^{2}_{1} \rangle$ and $\langle {\bf p}_1 \cdot {\bf p}_2 \rangle$ expectation values. Each of these
ways can be used in actual applications. For each of the helium isotopes considered in this study the expectation values of these three operators can be 
found in Tables I - III where all expectation values are given in atomic units. 

In addition to the mass-dependent components of the isotopic shift there is a component which is directly related to the proton density ditribution in the 
atomic nucleus. This is the field shift component of the isotopic shift. For light atoms (and ions) the overall contribution of this component is relatively 
small, but it plays an important role in some applications. In our calculations performed for this study we have used the Racah-Rosental-Breit formula 
(see, e.g., \cite{Sob} and references therein). In atomic units this formula takes the form 
\begin{eqnarray}
 \Delta E_{F} = \frac{4 \pi a^2_0}{Q} \cdot \frac{b + 1}{[\Gamma(2 b + 1)]^2} \cdot B(b) \alpha^{4 b} \cdot \Bigl( \frac{2 Q R}{a_0 \alpha^2} \Bigr)^{2 b} 
 \cdot \frac{\delta R}{R} \cdot \langle \delta({\bf r}_{eN}) \rangle \label{eqf3}
\end{eqnarray}
where $Q$ is the nuclear charge, $R$ is the nuclear radius, $a_0 \alpha^2 = r_e$ is the classical electron's radius and $b = \sqrt{1 - \alpha^2 Q^2}$, where 
$\alpha = \frac{e^2}{\hbar c} \approx \frac{1}{137}$ is the fine-structure (dimensionless) constant which is the small parameter in QED. In Eq.(\ref{eqf3}) 
the notation $\Gamma(x)$ stands for the Euler's gamma-function, while the factor $B(b)$ is directly related to the proton density distribution in the atomic 
nucleus. By assuming a uniform distribution of the proton density over the volume of the nucleus one finds the following expression for the factor $B(b)$ from 
Eq.(\ref{eqf3})
\begin{eqnarray}
 B(b) = \frac{3}{(2 b + 1) (2 b + 3)} \label{eqf4} 
\end{eqnarray}
For light nuclei with $Q \le 6$ we have $b \approx 1$ and $B \approx \frac15$. Such a choice corresponds to the uniform distribution of the proton charge
density over the whole volume of the nucleus. It is good approximation for all light nuclei. The formula, Eq.(\ref{eqf3}), has been used in many papers for 
numerical evaluations of the field component of the isotopic shift, or field shift, for short. In some works, however, this formula was written with a number 
of `obvious simplifications'. Many of such `simplifications' are based on the fact that for light nuclei the numerical value of the factor $b$ is close to 
unity. Furthermore, in some papers the factor $b$ was mistakenly called and considered as the Lorentz factor, while the actual Lorentz factor $\gamma$ 
is the inverse value of $b$, i.e., $\gamma = \frac{1}{b} = \frac{1}{\sqrt{1 - \alpha^2 Q^2}}$, which always exceeds unity. As follows from Eq.(\ref{eqf3})
in order to determine the field component of the isotopic shift in light atoms one needs to know the radius of the nucleus $R$ and the expectation value 
of the electron-nucleus delta-function $\langle \delta({\bf r}_{eN}) \rangle$. The nuclear radii of the ${}^{3}$He and ${}^{4}$He nuclei are 1.881 $fm$ 
and 1.672 $fm$, respectively \cite{CRC}. Here all radii of nuclei are given in $fermi$, where 1 $fm$ = 1 $\cdot 10^{-13}$ $cm$. The expectation values of 
the electron-nucleus delta-functions determined for the ground states in the ${}^{3}$He and ${}^{4}$He atoms allows one to evaluate the numerical values 
of the field shifts $\Delta E_{F}$ for the ground $1^1S(L = 0)-$states in the ${}^{3}$He and ${}^{4}$He atoms. In general, the expectation value of the 
electron-nucleus delta-function determines the electron density at distances $\approx \Lambda_e = \alpha a_0 = \frac{\hbar}{m_e c} \approx 3.862 \cdot 
10^{-11}$ $cm$ which is a `physical zero-distance' for the non-relativistic wave function(s). 
   
\section{Lowest-order QED correction}

Our non-relativistic approach used for numerical calculations of the lowest-order QED corrections is, in fact, the two-stage procedure. At the first stage 
we determine the lowest-order QED correction for the model helium atom with the infinitely heavy nucleus, i.e. for the ${}^{\infty}$He atom. The formula for 
the lowest-order QED correction $\Delta E^{QED}_{\infty}$ in the two-electron ion with infinitely heavy nucleus is written in the form (in atomic units) 
\begin{eqnarray}
 \Delta E^{QED}_{\infty} &=& \frac{8}{3} Q \alpha^3 \Bigl[ \frac{19}{30} - 2 \ln \alpha - \ln K_0 \Bigr] \langle \delta({\bf r}_{eN}) \rangle 
 + \alpha^3 \Bigl[ \frac{164}{15} + \frac{14}{3} \ln \alpha - \frac{10}{3} S(S + 1) \Bigr] \langle \delta({\bf r}_{ee}) \rangle \nonumber \\
 &-& \frac{7}{6 \pi} \alpha^3 \langle \frac{1}{r^{3}_{ee}} \rangle \label{eqf8}
\end{eqnarray}
where $\alpha = \frac{e^2}{\hbar c} = 7.2973525698 \cdot 10^{-3}$ is the fine structure constant (see above), $Q$ is the nuclear charge (in atomic units) 
and $S$ is the total electron spin. The ground states in all two-electron ions considered in this study are the singlet states with $S = 0$. Also, in this 
formula $\ln K_0$ is the Bethe logarithm (see, e.g., \cite{BS}, \cite{AB}). 

The last term in Eq.(\ref{eqf8}) is called the Araki-Sucher term, or Araki-Sucher correction, since this correction was obtained and investigated 
for the first time by Araki and Sucher \cite{Araki}, \cite{Such}. Note that the expectation value of the term $\langle \frac{1}{r^{3}_{ee}} \rangle$ is 
singular, i.e., it contains the regular (i.e. non-divergent) part and non-zero divergent part. General theory of the singular exponential integrals was 
developed in our earlier works (see, e.g., \cite{Fro2007} and references therein). In particular, in \cite{Fro2007} we have shown that the $\langle 
\frac{1}{r^{3}_{ee}} \rangle$ expectation value is determined by the following formula
\begin{eqnarray}
  \langle \frac{1}{r^{3}_{ee}} \rangle = \langle \frac{1}{r^{3}_{ee}} \rangle_R + 4 \pi \langle \delta({\bf r}_{ee}) \rangle \label{eqf85}
\end{eqnarray} 
where $\langle \frac{1}{r^{3}_{ee}} \rangle_R$ is the regular part of this expectation value and $\langle \delta({\bf r}_{ee}) \rangle$ is the expectation value 
of the electron-electron delta-function. The presence of non-zero divergent (or singular) parts in singular expectation values directly follows from the fact 
that the corresponding operators are self-conjugate. Here we cannot discuss this interesting, but non-trivial problem. Briefly, we can only say that the overall 
contribution of the singular part of the $\frac{1}{r^{3}_{ee}}$ operator is reduced to the expectation value of the electron-electron delta-function. Formally, 
in \cite{Fro2007} the equality, Eq.(\ref{eqf85}), was derived for the exponential variational expansion, Eq.(\ref{exp1}), only. However, it can be shown that the 
same equality is true in the general case. Analogous formula can be written for the electron-nucleus expectation value $\langle \frac{1}{r^{3}_{eN}} \rangle$. 
 
For the two-electron helium atoms with the finite nuclear masses we need to evaluate the corresponding recoil correction to the lowest-order QED 
correction. Such a correction is also given in \cite{Fro2007}. In atomic units it is written in the following form
\begin{eqnarray}
 \Delta E^{QED}_{M} &=& \Delta E^{QED}_{\infty} - \Bigl(\frac{2}{M} + \frac{1}{M + 1} \Bigr)
 \Delta E^{QED}_{\infty} + \frac{4 Q^2 \alpha^3}{3 M} \Bigl[ \frac{31}{3} + \frac{2}{Q} - \ln \alpha - 4 \ln K_0 \Bigr] \langle \delta({\bf r}_{eN}) 
 \rangle \nonumber \\
 &+& \frac{7 \alpha^3}{3 \pi M} \langle \frac{1}{r^{3}_{eN}} \rangle \label{eqf9}
\end{eqnarray} 
where $M \gg m_e$ is the nuclear mass. The difference $\Delta E^{QED}_{M} - \Delta E^{QED}_{\infty}$ is the recoil correction to the lowest order QED 
correction $\Delta E^{QED}_{\infty}$. In some works the recoil correction is defined as the absolute value of this difference. The inverse mass $\frac{1}{M}$ is a 
small parameter which is smaller than $\le 3 \cdot 10^{-4}$ (for all isotopes of the He atom). The dimensionless ratio $R = \frac{\mid \Delta E^{QED}_{\infty} 
- \Delta E^{QED}_{M} \mid}{\Delta E^{QED}_{\infty}}$ is small and can be evaluated as $R \approx \frac{1}{M} \ll 1$. To perform numerical calculations of 
$\Delta E^{QED}_{M}$ we used the nuclear masses of the ${}^{3}$He and ${}^{4}$He nuclei mentioned above. Note that all expectation values in Eq.(\ref{eqf9}) 
must be determined for actual two-electron ions, i.e. for ions with the finite nuclear masses. By using our expectation values of the electron-nucleus and 
electron-electron delta-functions and Araki-Sucher terms ($\langle \frac{1}{r^{3}_{ee}} \rangle$ and $\langle \frac{1}{r^{3}_{eN}} \rangle$) we have determined 
the lowest order QED corrections for each of the helium atom (${}^{\infty}$He, ${}^{4}$He and ${}^{3}$He) considered in this study. 

\section{Calculations and Conclusions}

First, let us determine the field components of the isotope shifts for the ${}^{4}$He and ${}^{3}$He atoms. By using Eq.(\ref{eqf3}) and the expectation values 
for the electron-nucleus delta-functions from Table IV we have found that $\Delta_F$(${}^{4}$He) $\approx$ 1.82052836$\cdot 10^{-8}$ $a.u.$ and 
$\Delta_F$(${}^{3}$He) $\approx$ 2.303733185$\cdot 10^{-8}$ $a.u.$ These values are the field components of the isotope shift determined for the ground 
$1^1S(L = 0)-$states in the two-electron ${}^{4}$He and ${}^{3}$He atoms. In these calculations we have used the following numerical factors in Eq.(\ref{eqf3}): $b 
\approx$ 0.999893491619, $2^{2b+2} \pi Q^{2b-1} \cdot \frac{b + 1}{[\Gamma(2 b + 1)]^2} \cdot B(b) \cdot \alpha^{4 b} \approx 2.8571973 \cdot 10^{-3}$ (for both 
isotopes), $\Bigl( \frac{R}{a_0 \alpha^2} \Bigr)^{2 b} = \Bigl( \frac{R}{r_e} \Bigr)^{2 b} \approx$ 0.3520929 (for the ${}^{4}$He atom) and $\Bigl( \frac{R}{a_0 
\alpha^2} \Bigr)^{2 b} = \Bigl( \frac{R}{r_e} \Bigr)^{2 b} \approx$ 0.4456063 (for the ${}^{3}$He atom).  

By using the expectation values of the operators from Table IV we determined the lowest order QED corrections for the ground $1^1S-$states of the ${}^{4}$He and 
${}^{3}$He atoms. In particular, for the model ${}^{\infty}$He atom the lowest order QED correction, i.e. the $\Delta E^{QED}_{\infty}$ value from Eq.(\ref{eqf8}), 
equals $\approx$ 2.226183190$\cdot 10^{-5}$ $a.u.$ The numerical value of the Bethe logarithm $\ln K_0$ = 4.3701602218 was taken from \cite{Fro2007}. From Eq.(\ref{eqf9}) 
we have found $\Delta E^{QED}_{M}({}^{4}$He$) \approx$ 2.225150320$\cdot 10^{-5}$ $a.u.$ and $\Delta E^{QED}_{M}({}^{3}$He$) \approx$ 2.224812380$\cdot 10^{-5}$ $a.u.$ In 
$Megahertz$ (or $MHz$) the corresponding corrections are: $\Delta E^{QED}_{\infty} \approx$ 1.464758174$\cdot 10^{5}$ $MHz$, $\Delta E^{QED}_{M}({}^{4}$He$) \approx$
1.464078578$\cdot 10^{5}$ $MHz$ and $\Delta E^{QED}_{M}({}^{3}$He$) \approx$ 1.463856224$\cdot 10^{5}$ $MHz$. Here we used the most recent conversion factor from $a.u.$ 
to $MHz$ which equals 6.579 683 920 729$\cdot 10^{9}$.    

We have performed highly accurate computations of the ground $1^1S-$states in the two-electron helium-3 and helium-4 atoms and in the model ${}^{\infty}$He atom. By using 
the computed expectation values of some operators we have evaluated (to high accuracy) all mass-dependent components and field component of the isotope shift(s) for the 
helium-3 and helium-4 atoms. We also evaluated the lowest-order QED corrections (or Quantun Elelctrodynamics corrections) for each isotope of the helium atom. Results of 
our study are of interest for future highly accurate calculations of the total isotopic shifts for different isotopes of the helium atom. In our analysis we derived and used 
in calculations the new (corrected) formula for the recoil correction to the lowest-order QED correction which can be applied to atoms/ions with arbitrary nuclear (electric) 
charge $Q \ge 1$. In earlier modifications of this formula, Eq.(\ref{eqf9}), the factor $Q^2$ in front of the third term was missing. The old formula was applicable to the 
negatively charged hydrogen ions (when $Q = 1$), but it was leading to certain contradictions for atomic systems with larger $Q$. In general, our results for the lowest-order 
QED corrections determined for these atomic systems coincide well with the corresponding results obtained in earlier studies (see, e.g., \cite{Dra08}, \cite{Eroh}). 
Nevertheless, quite a few modifications must be done in our procedure to improve the overall accuracy of our calculations of the lowest-order QED corrections. First of all, we 
need to improve our old approach which was derived and used 10 - 15 years ago to evaluate the Bethe logarithms for different two-electron atomic systems. Also, in future 
studies it will be very interesting to consider the lowest-order relativistic and QED corrections for other atomic systems and for different bound states in such systems. 

In conclusion we have to note that accurate numerical evaluations of the lowest-order QED corrections and other higher-order (upon $\alpha$) corrections are of increasing 
interest for various few-body atomic and molecular systems (see, e.g., \cite{Dra08}, \cite{Eroh}  and references therein for the two-electron helium atom(s) and \cite{H3+} for 
the H$^{+}_3$ ion). For two-electron helium atoms and helium-like ions this fact can be explained by a stream of experimental papers in which some new approaches to 
high-precision measurments have been developed and applied (see, e.g., \cite{Ref1}, \cite{Ref2} and \cite{Ref3}). A number of these new method are based on extensive use of 
lasers generating radiation at different frequencies (from infrared to vacuum ultraviolet regions). This allows one to determine the `absolute' positions (i.e. the total 
energies) of many atomic levels (or bound states) to extremely high accuracy which has been considered as `non-realistic' even fifteen years ago.

\newpage
 \begin{table}[tbp]
   \caption{The total energies $E$ and expectation values of some operators for the ground $1^1S(L = 0)-$state in the 
            two-electron ${}^{\infty}$He atom (in atomic units). $K$ is the total number of basis functions used.}
     \begin{center}
%     \scalebox{0.72}{%
     \begin{tabular}{| c | c | c | c |}
      \hline\hline
  $K$ & $E({}^{\infty}$He) & $\langle \delta({\bf r}_{eN}) \rangle$ & $\nu^{(a)}_{eN}$ \\
     \hline\hline
 3500 & -2.90372437703411959831028794 & 1.8104293184989490 & -2.0000000001645 \\
 
 3700 & -2.90372437703411959831041598 & 1.8104293184982854 & -2.0000000001566 \\

 3840 & -2.90372437703411959831052149 & 1.8104293185022350 & -2.0000000002048 \\

 4000 & -2.90372437703411959831060914 & 1.8104293185013928 & -2.0000000002296 \\
      \hline
 $K$ & $\frac12 \langle {\bf p}^2_1 \rangle$ & $\langle \delta({\bf r}_{ee}) \rangle$ & $\nu_{ee}$ \\
      \hline\hline
 3500 & 1.451862188517059799151774 & 0.106345370633423426 & 0.500000000385 \\

 3700 & 1.451862188517059799152265 & 0.106345370634289682 & 0.500000000295 \\

 3840 & 1.451862188517059799152708 & 0.106345370634901767 & 0.500000000167 \\

 4000 & 1.451862188517059799153096 & 0.106345370633985735 & 0.500000000219 \\
      \hline
 $K$ & $\langle {\bf p}_1 \cdot {\bf p}_2 \rangle$ &  $\langle (r^{-3}_{eN})_R \rangle$ & $\langle r^{-2}_{eN} \rangle$ \\
      \hline\hline
 3500 & 0.15906947508584375007426732 & -53.67642660233 & 6.0174088670242831 \\

 3700 & 0.15906947508584375007535182 & -53.67642660228 & 6.0174088670242472 \\

 3840 & 0.15906947508584375007642613 & -53.67642660269 & 6.0174088670242710 \\

 4000 & 0.15906947508584375007716180 & -53.67642660259 & 6.0174088670242867 \\
      \hline
 $K$ & $\frac12 \langle {\bf p}^2_N \rangle$ & $\langle (r^{-3}_{ee})_R \rangle$ & $\langle r^{-2}_{ee} \rangle$ \\
      \hline\hline
 3500 & 3.06279385211996334837781517 & -0.347101795395 & 1.4647709233190573 \\

 3700 & 3.06279385211996334837988169 & -0.347101795480 & 1.4647709233190648 \\

 3840 & 3.06279385211996334838183708 & -0.347101795529 & 1.4647709233190819 \\

 4000 &  3.0627938521199633483833533 & -0.347101795446 & 1.4647709233190672 \\
     \hline\hline
  \end{tabular}
  \end{center}
${}^{(a)}$The exact (or expected) value of the electron-nucleus cusp $\nu_{eN}$ in this case equals -2.0, while the exact (or 
expected) value of the electron-electron cusp $\nu_{ee}$ in this case equals 0.5.
  \end{table}

 \newpage
 \begin{table}[tbp]
   \caption{The total energies $E$ and expectation values of some operators for the ground $1^1S(L = 0)-$state in the 
            two-electron ${}^{4}$He atom (in atomic units). $K$ is the total number of basis functions used.}
     \begin{center}
%     \scalebox{0.72}{%
     \begin{tabular}{| c | c | c | c |}
      \hline\hline
  $K$ & $E({}^{4}$He) & $\langle \delta({\bf r}_{eN}) \rangle$ & $\nu^{(a)}_{eN}$ \\
     \hline\hline
 3500 & -2.90330455772956878574752636 & 1.8096724348881888 & -1.9997258506515 \\

 3700 & -2.90330455772956878574765410 & 1.8096724348875704 & -1.9997258507378 \\

 3840 & -2.90330455772956878574775934 & 1.8096724348914945 & -1.9997258507580 \\

 4000 & -2.90330455772956878574784673 & 1.8096724348906518 & -1.9997258507337 \\
      \hline
 $K$ & $\frac12 \langle {\bf p}^2_1 \rangle$ & $\langle \delta({\bf r}_{ee}) \rangle$ & $\nu_{ee}$ \\
      \hline\hline
 3500 & 1.451442403897224497512330 & 0.10629995671080796 & 0.500000000388 \\

 3700 & 1.451442403897224497512841 & 0.10629995671170843 & 0.500000000112 \\

 3840 & 1.451442403897224497513309 & 0.10629995671230651 & 0.500000000169 \\

 4000 & 1.451442403897224497513693 & 0.10629995671142032 & 0.500000000215 \\
      \hline
 $K$ & $\langle {\bf p}_1 \cdot {\bf p}_2 \rangle$ &  $\langle (r^{-3}_{eN})_R \rangle$ & $\langle r^{-2}_{eN} \rangle$ \\
      \hline\hline
 3500 & 0.15889694931179551018765512 & -53.65087859362 & 6.0157169308869035 \\

 3700 & 0.15889694931179551018882837 & -53.65087859357 & 6.0157169308868695 \\

 3840 & 0.15889694931179551019001349 & -53.65087859397 & 6.0157169308868947 \\

 4000 & 0.15889694931179551019069421 & -53.65087859387 & 6.0157169308869101 \\
      \hline
 $K$ & $\frac12 \langle {\bf p}^2_N \rangle$ & $\langle (r^{-3}_{ee})_R \rangle$ & $\langle r^{-2}_{ee} \rangle$ \\
      \hline\hline
 3500 & 3.06178175710624450521231532 & -0.346670375938 & 1.4643846304014152 \\

 3700 & 3.06178175710624450521450953 & -0.346670376026 & 1.4643846304014236 \\

 3840 & 3.06178175710624450521663053 & -0.346670376074 & 1.4643846304014405 \\

 4000 & 3.06178175710624450521807975 & -0.346670375994 & 1.4643846304014261 \\
     \hline\hline
  \end{tabular}
  \end{center}
${}^{(a)}$The exact (or expected) value of the electron-nucleus cusp $\nu_{eN}$ in this case equals -1.9997258508728739119366760743, while the exact (or 
expected) value of the electron-electron cusp $\nu_{ee}$ in this case equals 0.5.
  \end{table}

\newpage
 \begin{table}[tbp]
   \caption{The total energies $E$ and expectation values of some operators for the ground $1^1S(L = 0)-$state in the 
            two-electron ${}^{3}$He atom (in atomic units). $K$ is the total number of basis functions used.}
     \begin{center}
%     \scalebox{0.72}{%
     \begin{tabular}{| c | c | c | c |}
      \hline\hline
  $K$ & $E({}^{3}$He) & $\langle \delta({\bf r}_{eN}) \rangle$ & $\nu^{(a)}_{eN}$ \\
     \hline\hline
 3500 & -2.90316721071057772469861227 & 1.8094248565739434 & -1.9996361572278 \\  

 3700 & -2.90316721071057772469873991 & 1.8094248565743251 & -1.9996361579151 \\

 3840 & -2.90316721071057772469884505 & 1.8094248565752458 & -1.9996361585239 \\

 4000 & -2.90316721071057772469893238 & 1.8094248565764062 & -1.9996361596669 \\
      \hline
 $K$ & $\frac12 \langle {\bf p}^2_1 \rangle$ & $\langle \delta({\bf r}_{ee}) \rangle$ & $\nu_{ee}$ \\
      \hline\hline
 3500 & 1.451305083284658255476172 & 0.10628510207881447 & 0.500000000387 \\

 3700 & 1.451305083284658255476681 & 0.10628510207971530 & 0.500000000141 \\

 3840 & 1.451305083284658255477153 & 0.10628510208031233 & 0.500000000169 \\ 

 4000 & 1.451305083284658255477532 & 0.10628510207942670 & 0.500000000314 \\
      \hline
 $K$ & $\langle {\bf p}_1 \cdot {\bf p}_2 \rangle$ &  $\langle (r^{-3}_{eN})_R \rangle$ & $\langle r^{-2}_{eN} \rangle$ \\
      \hline\hline
 3500 & 0.15884052357094316093265572 & -53.64252204694 & 6.0151634482660996 \\

 3700 & 0.15884052357094316093382370 & -53.64252204689 & 6.0151634482660655 \\

 3840 & 0.15884052357094316093500807 & -53.64252204729 & 6.0151634482660907 \\

 4000 & 0.15884052357094316093568536 & -53.64252204719 & 6.0151634482661061 \\
      \hline
 $K$ & $\frac12 \langle {\bf p}^2_N \rangle$ & $\langle (r^{-3}_{ee})_R \rangle$ & $\langle r^{-2}_{ee} \rangle$ \\
      \hline\hline
 3500 & 3.06145069014025967188499925 & -0.346529284097 & 1.4642582634664809 \\

 3700 & 3.06145069014025967188718641 & -0.346529284186 & 1.4642582634664893 \\

 3840 & 3.06145069014025967188930618 & -0.346529284233 & 1.4642582634665062 \\

 4000 & 3.06145069014025967189075025 & -0.346529284153 & 1.4642582634664919 \\
     \hline\hline
  \end{tabular}
  \end{center}
${}^{(a)}$The exact (or expected) value of the electron-nucleus cusp $\nu_{eN}$ in this case equals -1.9996361575870467745067283848, while the exact (or 
expected) value of the electron-electron cusp $\nu_{ee}$ in this case equals 0.5.
  \end{table}

 \newpage
 \begin{table}[tbp]
   \caption{The expectation values of the delta-functions and other operators used in calculations of the 
            $\Delta E^{QED}_{\infty}$ and $\Delta E^{QED}_{M}$ corrections for the ground $1^1S(L = 0)-$state in 
            the two-electron ${}^{4}$He atom (in atomic units).}
     \begin{center}
%     \scalebox{0.72}{%
     \begin{tabular}{| c | c | c | c | c |}
      \hline\hline
          &  ${}^{\infty}$He   &   &  ${}^{4}$He    &     ${}^{3}$He        \\
     \hline\hline
 $\langle \delta({\bf r}_{eN}) \rangle$ & 1.810429318501 & $\langle \delta({\bf r}_{eN}) \rangle$ &  1.809672434890 & 1.8094248565750 \\ 
           \hline
 $\langle (r^{-3}_{ee})_R \rangle$ & -0.347101795480     &  $\langle (r^{-3}_{eN})_R \rangle$     &  -53.65087859371 & -53.64252204710 \\
            \hline
 $\langle \delta({\bf r}_{ee}) \rangle$ & 0.106345370634 &  ----------------    &  -------------- & --------------- \\                                    
     \hline\hline
  \end{tabular}
  \end{center}
  \end{table}

\begin{thebibliography}{10}

\bibitem{FrFi} C. Froese Fisher, T. Brage and P. J\"{o}nsson, {\it Computational Atomic Structure}, (IOP, Bristol (UK) (1997)).

\bibitem{CRC} \textit{CRC Handbook of Chemistry and Physics}, 92nd Edition, Ed. W.M. Haynes, (Taylor and Francis Group, Boca Raton, FL, (2011-2012)).

\bibitem{Fro98} A.M. Frolov, Phys. Rev. A {\bf 57}, 2436 (1998).

\bibitem{MQW} R. McWeeny and B.T. Sutcliffe, {\it Methods of Molecular Quantum Mechanics}, (Academic Press, New York, (1969)), Chps. 4 and 5.

\bibitem{Sob} I.I. Sobelman, {\it Atomic Spectra and Radiative Transitions}, (Spinger, Berlin (1979)).

\bibitem{BS} H.A. Bethe and E.E. Salpeter, {\it Quantum Mechanics of One- and Two-Electron Atoms}, (Dover Publ. Inc., Mineola, NY, (2008)).

\bibitem{AB} A.I. Akhiezer and V.B. Beresteskii, {\it Quantum Electrodynamics}, (4th ed., Science, Moscow (1981)), Chps. 4 and 5.

\bibitem{Araki} H. Araki, Prog. Theor. Phys. {\bf 17}, 619 (1957).

\bibitem{Such} J. Sucher, Phys. Rev. {\bf 109}, 1010 (1958).

\bibitem{Fro2007} A.M. Frolov, J. Chem. Phys. {\bf 126}, 104302 (2007).

\bibitem{Dra08} G.W.F. Drake and Z.C. Yan, Can. J. Phys. {\bf 86}, 45 (2008).

\bibitem{Eroh} V.A. Yerokhin and K. Pachucki, Phys. Rev. A {\bf 84}, 062512 (2010).

\bibitem{H3+} L. Lodi, O. Polyansky, J. Tennyson et al., Phys. Rev. A {\bf 89}, 032505 (2014).

\bibitem{Ref1} P.L. Luo, J.L. Peng, J.T. Shy et al., Phys. Rev. Lett. {\bf 111}, 013002 (2013).

\bibitem{Ref2} P.L. Luo, J.L. Peng, J.T. Shy et al., Phys. Rev. Lett. {\bf 111}, 179901 (2013).

\bibitem{Ref3} D.Z. Kandula, C. Gohle, T.J. Pinkert et al., Phys. Rev. A {\bf 84}, 062512 (2011).

\end{thebibliography}
\end{document}